\documentclass[showpacs,twocolumn,floatfix]{revtex4-1}
\usepackage{graphicx,psfrag,amsmath,amssymb,amsfonts,latexsym,color,dcolumn,
epsf,graphpap}

\begin{document}

\title{Nonunitary geometric phases: a qubit coupled to 
an environment with random noise}

\author{Fernando C. Lombardo\footnote{lombardo@df.uba.ar}
and Paula I. Villar\footnote{paula@df.uba.ar}}
\affiliation{Departamento de F\'\i sica {\it Juan Jos\'e
Giambiagi}, FCEyN UBA and IFIBA CONICET-UBA, Facultad de Ciencias Exactas y Naturales,
Ciudad Universitaria, Pabell\' on I, 1428 Buenos Aires, Argentina}

\date{today}

\begin{abstract}
We describe the decoherence process induced on a two-level quantum system
in direct interaction with a non-equilibrium environment. 
The non-equilibrium feature is represented
by a non-stationary random function corresponding to the fluctuating transition
frequency between two quantum states coupled to the surroundings.
In this framework, we compute the decoherence factors which have a characteristic
``dip'' related to the initial phases of the bath modes. We therefore study
different types of environments, namely ohmic and supra-ohmic. These environments
present different decoherence time-scales than the thermal environment we used
to study. As a consequence, we compute analytically and numerically the non-unitary geometric phase
for the qubit in a quasi-cyclic evolution under the presence of these
particular non-equilibrium environments. We show in which cases decoherence effects 
could, in principle, be controlled in order to perform a measurement of the geometric 
phase using standard procedures.

\end{abstract}

\pacs{05.40.-a;05.40.Ca;03.65.Yz}

\maketitle

\newcommand{\beq}{\begin{equation}}
\newcommand{\eeq}{\end{equation}}
\newcommand{\dalam}{\nabla^2-\partial_t^2}
\newcommand{\mbf}{\mathbf}
\newcommand{\itm}{\mathit}
\newcommand{\beqa}{\begin{eqnarray}}
\newcommand{\eeqa}{\end{eqnarray}}

\section{Introduction}
\label{intro}

The spin-boson model is studied in a variety of fields, such as
condensed matter physics, quantum optics, physical chemistry and
quantum information science \cite{legget} in order to describe
non-unitary effects induced in quantum systems due to a
coupling with an external environment.
For a quantum system, the influence of the surroundings 
plays a role at a fundamental level. When the environment 
is taken into consideration, the system dynamics can no 
longer be described in terms of pure quantum
states and unitary evolution. From a practical point of view,
all real systems interact with an environment to a greater or lesser extent,
which means that we expect their quantum evolution to be plagued by non-unitary
effects, namely dissipation and decoherence.

Most theoretical investigations of how the system is affected
by the presence of an environment have been done using a thermal reservoir,
usually assuming Markovian statistical properties and defined bath correlations 
\cite{paz-zurek, weiss} (there are also works on non-Markovian models as, just for example, \cite{nonMark}). 
However, there has been some growing interest in modelling
more realistic environments, sometimes called ``composite'' environments \cite{lutz,pracomp,pra3}.
In fact, the are many situations where the environment is better modelled 
by a non-equilibrium bath.  Quantum  dynamics in non-equilibrium environments has been previously
considered by some recent investigations. For example, light-induced ultra-fast coherent
electronic processes in chemical or biological systems may occur
on sufficiently short time scales \cite{martensiop}. In these cases, initial non-equilibrium
states induced in the bath through the coupling among system and environment,
might not have the chance to reach equilibrium rapidly. Then, the transient
non-equilibrium bath dynamics may undergo a non-trivial interaction with the system
of interest in comparable time scales.
 Gordon $\it {et~ al}$ discussed
 the control of quantum coherence and the suppression of dephasing by 
 stochastic control fields \cite{Gordon}. In \cite{lutz},
 the decoherence process induced by a non-equilibrium environment described
 by several equilibrium baths at different temperatures, is discussed. Therein, it was 
suggested that the effect of such environment on the quantum system could be 
 described as the effect done by a single effective bath with a time-dependent
 temperature. The decoherence of single trapped ions coupled to engineered
 reservoirs, where the internal state and coupling can be controlled was studied in \cite{Myatt}.

In this context, we shall describe a simple model which gives a different insight into the behaviour of a 
quantum system coupled to an environment that is not at thermal
equilibrium. Herein, we study the dynamics of quantum coherence in non-equilibrium. 
We consider a two level quantum system in a non-equilibrium bath, represented
by random perturbations with non-stationary statistics. Therefore, we shall study
how the quantum system is affected by the decoherence induced by the environment.
We shall compare this decoherence process with the usual results for a thermal environment.

From another point of view, a system can retain the information of
its motion when it undergoes a cyclic evolution in the form of a
geometric phase (GP), which was first put forward by Pancharatman in
optics \cite{Pancharatman} and later studied explicitly by Berry in
a general quantal system \cite{Berry}. Since then, great progress
has been achieved in this field.  As
an important evolvement, the application of the geometric phase has
been proposed in many fields, such as the geometric quantum
computation. Due to its global properties, the geometric phase is
propitious to  construct  fault tolerant  quantum  gates. In this
line of work, many physical systems  have  been  investigated  to
realize  geometric  quantum  computation,  such  as  NMR  (Nuclear
Magnetic Resonance) \cite{NMR} , Josephson  junction \cite{JJ},  Ion  trap \cite{IT} and
semiconductor  quantum  dots \cite{QD}. The quantum computation scheme for the 
GP has been proposed based on the Abelian  or
non-Abelian geometric concepts, and the GP has been shown
to be robust against faults in the presence of some kind of
external noise due to the geometric nature of Berry phase \cite{refs1, refs2, refs3}. It was
therefore seen that the interactions play an important role for the
realization of some specific operations. As the gates operate slowly compared to the 
dynamical time scale, they become vulnerable to open system effects and parameters fluctuations
that may lead to a loss of coherence.
Consequently, study of the
GP was soon extended to open quantum systems. Following this idea,
 many authors have analysed the correction to the GP
 under the influence of an external thermal environment using different
approaches (see \cite{Tong, pra, nos, pau} and references therein).

In this paper, we shall study how the GP is affected by the presence of a 
non-equilibrium environment. We shall consider a two-state quantum system 
coupled to such an environment and derive the corresponding decoherence
factor in Sec.\ref{Deco}. We shall analyse the decoherence process for ohmic
and non-ohmic environments. In Sec.\ref{GPs}, we shall derive the GP for a
non-unitary evolution of the quantum system in the presence of a non-equilibrium 
environment and compute how the GP is corrected in each case. Finally, in
Sec.\ref{conclu}, we shall make our final remarks.


\section{Purely dephasing solvable spin-boson model}
\label{Deco}

A paradigmatic model of open quantum systems is a two-state quantum system coupled to
a thermal environment. This is a particular case of
the spin-boson model by A. Leggett \cite{legget} (where the tunnelling
bare matrix element is $\Delta=0$) and has been used by many
authors to model decoherence in quantum computers \cite{Ekert}
and, in particular, it is extremely relevant to the proposal 
for observing GPs in a superconducting nano-circuit \cite{Falci}.
In spite of its simplicity, this model captures many of the
elements of decoherence theories and sheds some insight into the
modification of the GPs due to the presence of the environment. 
The interaction between the two-state system and the environment
is entirely represented by a Hamiltonian in which the coupling is
only through $\sigma_z$. In this particular case,
$[\sigma_z,H_{\rm int}]=0$ and the corresponding master equation for the reduced density matrix, is
much simplified, with no frequency renormalization and dissipation
effects. In other words, the model describes a purely decohering (dephasing) mechanism,
solely containing the diffusion term ${\cal D}(t)$ \cite{pra}. In such a case, it is easy to
check that
 \begin{eqnarray}
\rho_{\rm r_{01}}(t) &=& e^{-i \Omega t} e^{-{\cal A}(t)}
~\rho_{\rm r_{01}}(0), \nonumber
\end{eqnarray}
is the solution for the off-diagonal terms of the reduced density matrix
(while populations remain constant) and ${\cal A}= \int_0^{\infty} ds {\cal D}(s)$.
$\Omega$ refers to the angular frequency of precession of a spin precessing the
$z$ axis as ruled by the isolated from the environment Hamiltonian $H_0= \frac{1}{2} \Omega \sigma_z$ 
(responsible for the unitary evolution). The spin-boson model is the one used in Ref.\cite{pra} in order to present a solvable 
model to study how the GPs are corrected by decoherence in open systems. In that framework, 
we have studied not only how the GPs are corrected by the presence of the different type of 
environments but estimated the corresponding times at which decoherence become effective as well. These 
estimations should be taken into account when planning experimental setups, as the one performed in Ref. \cite{prl}, where using a NMR 
quantum simulator, the geometric phase of an open system undergoing nonunitary evolution han been obtained. The GP was computed 
in a tomographic manner, measuring the off-diagonal elements of the reduced density matrix of the system. This   
study of the GP  in the nonunitary regime is particularly important for the application of fault-tolerant quantum computation (see \cite{scienceexp} as an example of measuring the Berry phase in a solid-state qubit where there is an important geometric contribution to dephasing that occurs when geometric operations are carried out in the presence of low-frequency noise). 

In the present paper, we shall adopt a different model of decoherence than the one in \cite{pra}. We are 
concerned with non-equilibrium situations, in which the qubit (the main quantum system) is coupled to 
a non-equilibrium bath. The two-level quantum system presents an energy gap 
$E_2(t)-E_1(t)= \hbar \omega(t)$ which fluctuates due to the influence of the environment,
where $E_j(t)$, with $j=1,2$ is the instantaneous energy of state $j$ as perturbed
by the surroundings. Following the idea
proposed in \cite{martenschem}, the bath is represented by a random function
of time corresponding to the transition frequency of the two-state quantum
system. In contrast to the usual treatment, the statistical properties
of this random function are non-stationary, corresponding physically to, for example,
impulsively excited phonons of the environment with initial phases that are not random,
but which have defined values at $t=0$. Due to this assumptions, this
environment is not at thermal equilibrium. The time-dependent frequency is written
in the form $ \omega (t) = \Omega + \delta \omega(t)$, where $\delta\omega(t)$ is defined as
\begin{equation}
  \delta \omega (t) = \sum_{k=1}^{\infty} c_k \cos(\omega_k t + \theta_k(t)).
\end{equation}

The Fourier components $c_k$ are positive constants related to the spectral 
density of the environment and the coupling of the bath modes to the system.
It is important to mention, that, in this model, the randomness enters through the
non-stationary distribution of random phases $\theta_k(t)$, which are given 
by $\theta_k(t)= \theta_k(0) + x_k(t)$. The random function $x_k(t)$ satisfies
a diffusion equation
\begin{equation}
 \partial_t P_k(x,t) = D_k \partial_x^2 P_k(x,t)
 \label{difeq}
\end{equation}
where $P_k(x,t)$ is a time-dependent probability distribution and
$D_k$ is the diffusion constant. The quantity $x_k$ is an angle, 
so $P(x,t)$ is a function with period $2 \pi$.
The time-dependent probability distribution for component $k$ that solves
Eq.(\ref{difeq}) with an initial localized condition $P(x,0)=\delta(x)$
is
\begin{equation}
 P_k(x,t)=\frac{1}{2 \pi} + \frac{1}{\pi} \sum_{n=1}^{\infty} e^{-n^2 D_k t}
 \cos(n x).
\end{equation}
This means that, physically, the phase of each component of the random force
is not random at $t=0$, when an impulsive excitation creates a quantum coherence
in the system, but decays to an uniform $1/2 \pi$ distribution under diffusive 
evolution with diffusion constant $D_k$ \cite{martenschem}.

Following this approach, it is possible to evaluate, 
\begin{eqnarray}
\rho_{\rm r_{01}}(t) &=& e^{-i \Omega t} < e^{- i \int_0^t \delta \omega(s)ds}> 
\rho_{\rm r_{01}}(0) \nonumber \\
&\equiv& e^{-i \Omega t} {\cal F}(t) ~\rho_{\rm r_{01}}(0),
\end{eqnarray} the solution for the off-diagonal element of the density matrix (while
the populations remain constant again). Here, we denote with $<...>$ the non-equilibrium average
over the non-stationary random bath and ${\cal F}(t)$ is defined as the decoherence
factor.

By considering the typical factor $f_k(t)= \exp(-i \int_0^t \delta \omega_k(s)ds)$,
one has to do some algebra to obtain $|{\cal F}(t)|$. This mainly consists on performing the time
integral and the averaging $f_k (t)$ over the distribution probability $P_k(x_k,t)$ \cite{martenschem}.
After these computations, a simple but accurate approximation can be obtained, namely
\begin{equation}
|{\cal F}(t)| =|\prod_{k} f_k (t)|\simeq e^{-\beta(t)},
\end{equation}
with
\begin{eqnarray}
 \beta(t) &=& \frac{1}{4} \int_0^{\infty} d\omega I(\omega)  \label{beta} \\
 & \times & \biggl[1-e^{-2 D t}+ (e^{-2 D t}-e^{-4 D t}) \cos(2(\omega t + \theta(\omega)))
 \biggr].
 \nonumber
\end{eqnarray}
It is important to note that in Eq.(\ref{beta}), the continuum limit has already
been taken (in the number of bath modes) and the diffusion constant has been assumed
$D(\omega)= D$ for simplicity.

In order to study the decoherence process induced in the system by the presence
of a non-equilibrium environment, we define a widely used physical spectral
density $I(\omega)= 4 \gamma/\Lambda^2 \omega^n/\Lambda^{n-1} e^{-\omega/\Lambda}
$ \cite{legget}, where $\gamma$ is the dimensionless dissipative constant
 and $\Lambda$ is the cutoff frequency. On general grounds, $\Lambda$ is the biggest frequency present in the 
environment, i.e. the frequency range of the environmental modes. In particular, the case with
$n=1$ is the ``ohmic" case and the one with $n>1$ is the
``supraohmic" case. The ohmic environment is the most studied case in the literature, for example in the quantum Brownian motion paradigm, and 
produces a dissipative force that in the limit of the frequency cutoff $\Lambda \rightarrow 0$ is proportional to the velocity. The 
supraohmic case, on the one hand, is generally used to model the interaction between defects and phonons in 
metals \cite{legget} and also to mimic the interaction between a charge and its own electromagnetic field (see for example \cite{sonnen}). In particular, the use of the supra-ohmic case can be used as a toy model to study decoherence process in quantum field theory \cite{qft}.  

This model of non-equilibrium is characterized by a key quantity which considers 
the effect of the initial phases of the bath modes in the
function $\theta(\omega)$. In this case, we consider a linear
dependence such as $\theta(\omega)=-\lambda \omega$. It is interesting to have the possibility to 
control dephasing by varying the single parameter $\lambda$. Following Eq.(\ref{beta}), the decoherence factors can be exactly calculated and they are given by 
\begin{equation}
 {\cal F}_{\rm ohmic} = e^{-\gamma e^{-4 D t} (-1 + e^{2 D t}) \left[  e^{2 D t} + \frac{1 - 4 \Lambda^2 (t - \lambda)^2 }{(1 + 4 \Lambda^2 (t - \lambda)^2)^2}  \right]},
        \label{fn1}
\end{equation}
for the ohmic case, and 
\begin{equation}
 {\cal F}_{\rm supra} = e^{-6 \gamma ~e^{-4 D t} (-1 + e^{2 D t})  \left[e^{2 D t} + \frac{
    1 - 24 (t - \lambda)^2 \Lambda^2 + 
     16 (t - \lambda)^4 \Lambda^4}{(1 + 
      4 (t - \lambda)^2 \Lambda^2)^4}\right]}
      \label{fn3}
\end{equation}
for the supraohmic case (we use $n=3$ all along this article).

It is interesting to analyse the asymptotic behaviour of the function $\beta(t)$. Both types 
of non-equilibrium environments produce a linear time-dependence for the very short time-scale $D t$, $\Lambda t \ll 1$, which 
induces a decoherence factor of the form ${\cal F} \sim \exp[- a \gamma t]$ (where $a$ is a constant with proper units). This is similar to the 
decoherence factor calculated in Ref. \cite{pra} for the case of an ohmic finite temperature 
environment (just assuming that $\gamma \sim \gamma_0 K_B T$).  In this limit, decoherence is always an efficient process, unless the dissipative constant $\gamma$ is very small. In the long time limit $D t$, $\Lambda t \gg 1$, both $\beta$ functions (the 
one for the ohmic, and the corresponding to the supraohmic cases) acquire a constant value (different for each type of environment). In this 
long time regime, decoherence factor in the ohmic case behaves as ${\cal F} \sim \exp[- \gamma ]$, similar to the decoherence 
factor for the equilibrium supraohmic environment at zero temperature \cite{pra}. Meanwhile, in the supraohmic case, the decoherence factor 
approaches a long time value given by ${\cal F} \sim \exp[-6 \gamma ]$. Again, as we mentioned before, 
in the case of small $\gamma$, decoherence never occurs, even at very long times. Intermediate times are 
ruled by the specific randomness introduced into the model. All in all, it is important to note the richness of the model which guarantees known and unknown decoherence processes by the correct tuning of the parameters.

\begin{figure}[!ht]
\includegraphics[width=9cm]{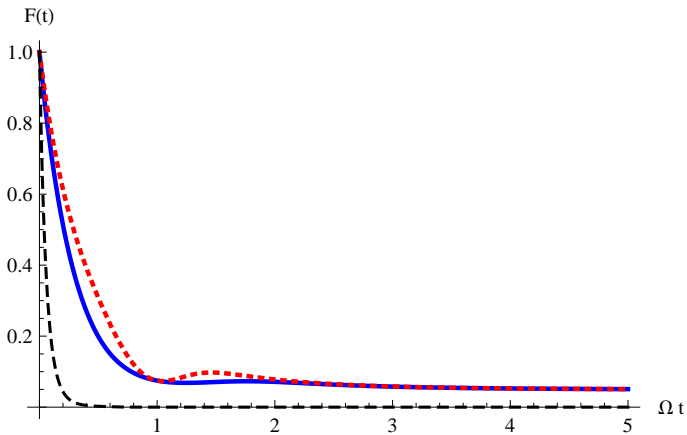}
\caption{(Color online) Evolution in time of the decoherence factor ${\cal F}(t)$ for different
models of the environment in the strong coupling limit. The red short-dashed line is for an ohmic non-equilibrium
bath and the long-dashed black line is for the supraohmic. For $D t$, $\Lambda t \gg 1$, the decoherence factor behaves as 
${\cal F}_{\rm ohmic}(t) \sim \exp(-\gamma)$
and ${\cal F}_{\rm supra}(t) \sim \exp(-6 \gamma)$.  We also present a solid blue line for the assumptions made
in \cite{martensiop}, a non-equilibrium bath with a Gaussian spectral density. 
Parameters used: $\gamma = 3$; $\Lambda/\Omega= 1$; $\Omega\lambda = 1$; $D/\Omega = 0.5$; and $n=3$ for the 
supraohmic environment.}
\label{Fggrande}
\end{figure}

In Fig. \ref{Fggrande}, we present the behaviour of the decoherence
factor for a strong dissipative case for both environments. As expected,
the decoherence factor decays from unity to an asymptotic value.
Therein, we also present the behaviour of the decoherence factor found
in \cite{martensiop}, where a different spectral
density to describe the environment has been used. The parameters used in the Figure are similar
to those used in \cite{martensiop} in order to do a better comparison and analysis. Unlike typical studies 
using the master equation in the weak coupling limit ($\gamma \sim c_k^2$), 
in the present approach there is no constraint for the value of $\gamma$. Therefore we can use either a
strong or weak coupling as a value for $\gamma$.
In the Figure, we can note three different
lines: the solid blue line for the results in \cite{martensiop}, the red short-dashed line for
our ohmic environment and the solid black long-dashed line for the supra-ohmic environment. Then, it
is easy to see that the ohmic case is very similar to the one obtained in \cite{martensiop},
where a Gaussian spectral density was considered. In both cases, decoherence is very efficient, as 
expected since we are considering the overdamped case ($\gamma \geq 1$). 
There is also an interesting fact: the supraohmic decoherence
factor has a smaller decoherence time-scale than the other two decoherence factors 
herein considered. This is unlike the case for equilibrium supraohmic environment, where 
decoherence is effective only at high temperature. 
This modelling of the environment gives a decoherence factor which drops from its initial
value toward an asymptotic value (${\cal F}(t \rightarrow \infty)$) 
after the intermediate time $t=\lambda$. At this time the system re-phases back 
to the slowly decaying envelope not purely exponential. As $\lambda$ becomes
large and positive, the decay approaches the envelope function without the non-monotonic dip (that occurs at $t = \lambda$).
Non-exponential behaviour in the decay of quantum coherence has been observed in full many-particle
simulations of quantum coherent dynamics under non-equilibrium conditions \cite{riga}.

\begin{figure}[!ht]
\includegraphics[width=8cm]{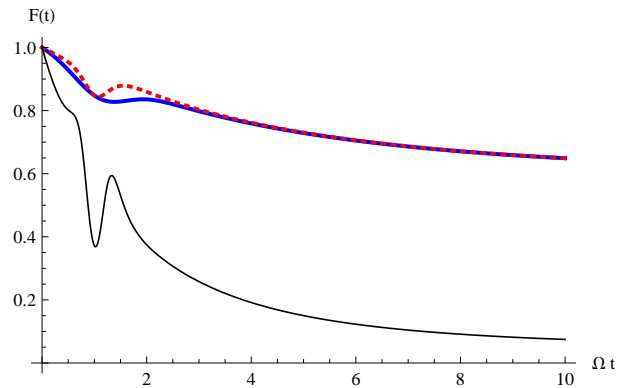}
\caption{(Color online) Evolution in time of the decoherence factor $F(t)$ for different
models of the environment in the weak coupling limit. The red short-dashed line is for an ohmic non-equilibrium
bath and the black solid line is for the supraohmic case. We also present a solid blue line for the assumptions made
in \cite{martensiop}, a non-equilibrium bath with a Gaussian spectral density (different with respect to the 
normally used in the theory of quantum open systems theory).
Parameters used: $\gamma = 0.5$; $\Lambda/\Omega= 1$; $\Omega\lambda = 1$; $D/\Omega = 0.1$.}
\label{Fgchico}
\end{figure}

\begin{figure}[!ht]
\includegraphics[width=8cm]{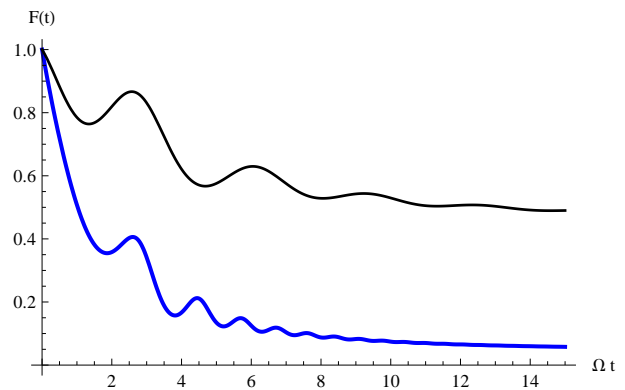}
\caption{(Color online) A more complex modelling of the initial phases of the bath modes by considering $\theta(\omega)=-\lambda \omega^2$ in an ohmic (blue solid line) and supraohmic (black solid line) environment.
Parameters used: $\gamma = 3$; $\Lambda/\Omega= 1$; $\Omega\lambda = 1$; $D/\Omega = 0.1$.}
\label{Fgamagc}
\end{figure}

In order to have a better view of the dip where the system ``recoheres" for a while, we present
the behaviour of the decoherence factors for the weak coupling case in Fig.\ref{Fgchico}.
Therein, $\gamma$ has a smaller value, comparable to those we used when dealing with
environments in thermal equilibrium in the underdamped case \cite{pla1,pla2, pra}.
 The dip is obtained by introducing, as we mentioned before, the 
 simple relation $\theta (\omega) =- \lambda \omega $ for the initial phases of the bath modes
in the modelling of the environment. Even though this assumption is a minimalistic model, it allows
to have some kind of control in the decoherence process which in turn can be useful in
experimental setups where decoherence is always an obstacle to overcome. This
result agrees with the one in \cite{lutz}, where it was shown that non-equilibrium decoherence can be slowed 
down in a controlled manner as compared to the corresponding equilibrium situation.

A different modelling of the initial phases of the bath modes can, 
in principle, be adopted. However, herein we use the linear one just for simplicity. A complex assumption can be, for example, $\theta(\omega)=-\lambda \omega^2$. The derivation of the decoherence factor is somewhat more difficult and is not worth writing explicitly here. Anyway, the decoherence factor for a quadratic behaviour in $\omega$ is presented in Fig. \ref{Fgamagc} for an ohmic and supra-ohmic non-equilibrium environment. In such a case, it is important 
to note a more complicated structure of dips in the decoherence factor.

\section{Application: Geometric phase for a qubit coupled to a non-equilibrium environment}
\label{GPs}

In order to compute the GP and note how it is corrected by the environment, we shall 
 briefly review  the way the
geometric phase can be computed for a system under
the influence of external conditions such as an external bath.
 In Ref. \cite{Tong}, a quantum kinematic
approach was proposed and the geometric phase
(GP) for a mixed state
under non-unitary evolution has been defined
as
\begin{eqnarray} \phi_G & = &
{\rm arg}\bigg\{\sum_k \sqrt{ \varepsilon_k (0) \varepsilon_k (\tau)}
\langle\Psi_k(0)|\Psi_k(\tau)\rangle \times \nonumber \\
& & e^{-\int_0^{\tau} dt \langle\Psi_k|
\frac{\partial}{\partial t}| {\Psi_k}\rangle}\bigg\}, \label{fasegeo}
\end{eqnarray}
where $\varepsilon_k(t)$ are the eigenvalues and
 $|\Psi_k\rangle$ the eigenstates of the reduced density matrix
$\rho_{\rm r}$ (obtained after tracing over the reservoir degrees
 of freedom). In the last definition, $\tau$ denotes a time
after the total system completes a cyclic evolution when it is
isolated from the environment. Taking into account the effect of the
environment, the system no longer undergoes a cyclic evolution.
However, we shall consider a quasi cyclic path ${\cal P}:~ t~
\epsilon~ [0,\tau]$, with $\tau= 2 \pi/ \Omega$ ($\Omega$ is the
system's characteristic frequency). When the system is open, the
original GP, i.e. the one that would have been obtained if the
system had been closed $\phi_U$, is modified. This means, in a
general case, that the phase can be interpreted as $\phi_G = \phi_U + \delta \phi_G$,
where $\delta \phi_G$ depends on the kind of environment coupled to
the main system \cite{pra, nos, pau, pra2, prl, pra3}.

Assuming an initial quantum state of the form
\begin{equation}
 |\psi (0) > = \cos(\frac{\theta_0}{2}) |0> + ~\sin(\frac{\theta_0}{2}) |1>,
\end{equation}
its evolution at a later time $t$, is
\begin{equation}
 |\psi (t) > = e^{-i \Omega t} \cos(\theta_+) |0> + ~\sin(\theta_+) |1>,
\end{equation}
where 
\begin{equation}
 \cos(\theta_+) = \frac{ \sin(\theta_0) |{\cal F}(t)|}{\sqrt{\sin^2( \theta_0)|{\cal F}(t)|^2 +
 4(\varepsilon_+ - \cos^2(\frac{\theta_0}{2}))^2}},
\end{equation}
\begin{equation}
 \sin(\theta_+) = \frac{2(\varepsilon_+ - \cos^2(\frac{\theta_0}{2})) }{\sqrt{\sin^2( \theta_0)|{\cal F}(t)|^2+
 4(\varepsilon_+ - \cos^2(\frac{\theta_0}{2}))^2}},
\end{equation}
and $\varepsilon_+$ the eigenvalue of the reduced density matrix, namely
\begin{equation}
 \varepsilon_+ = \frac{1}{2} \biggl( 1 + \sqrt{\cos^2(\theta_0) + \sin^2(\theta_0) |{\cal F}(t)|^2}
\biggr),
\end{equation}
while $\varepsilon_-$ does not contribute to the geometric case since $\varepsilon_-(t=0)=0$ (see definition Eq.(\ref{fasegeo})).

As in our previous works \cite{pra,pra2,pra3,prl},
the GP is obtained by computing eigenvectors and eigenvalues of the
reduced density matrix and using Eq.(\ref{fasegeo}),
\begin{equation}
 \phi_G = \Omega \int_0^{\tau} ~\cos^2(\theta_+)~dt.
 \label{fasecos}
\end{equation}

\begin{figure}[!ht]
\includegraphics[width=10cm]{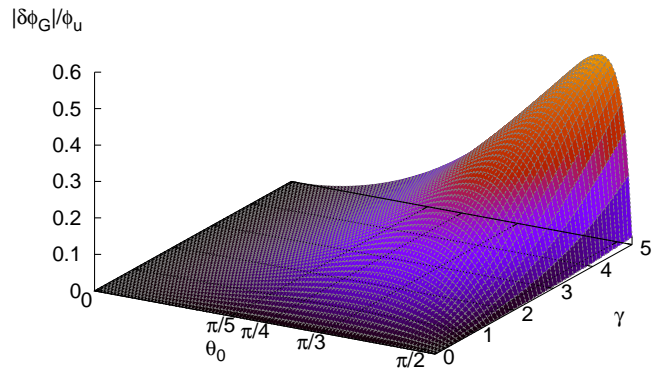}
\caption{(Color online) Behaviour of the geometric phase as a function of the initial state of the
quantum system $\theta_0$ (measured in radians) and the dissipation of the environment (dimensionless $\gamma$) induced by an ohmic non-equilibrium
environment in a cycle.
Parameters used:  $\Lambda/\Omega= 1$; $\Omega \lambda = 1$; $D/\Omega = 0.1$.}
\label{Fase3dn1}
\end{figure}

\begin{figure}[!ht]
\includegraphics[width=10cm]{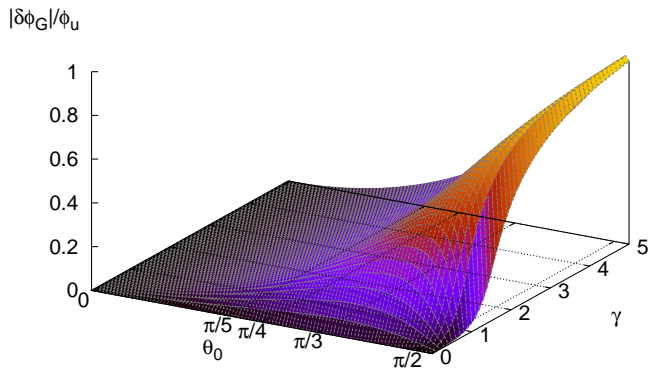}
\caption{(Color online) Behaviour of the GP as a function of the initial state ($\theta_0$ in radians) of the
quantum system and the dissipation of the environment (dimensionless $\gamma$) induced by a supraohmic non-equilibrium
environment in a cycle.
Parameters used:  $\Lambda/\Omega= 1$; $\Omega \lambda = 1$; $D/\Omega = 0.1$.}
\label{Fase3dn3}
\end{figure}

In Figs. \ref{Fase3dn1} and \ref{Fase3dn3} we plot the environmentally induced correction to the unitary phase, $\vert \delta \phi_G\vert$ (normalized by the value of $\phi_U$) as a function of the system's initial quantum state ($\theta_0$) and the
dissipation induced in the quantum subsystem due to the presence of the random environment $(\gamma)$.
In both Figures we have considered a wide range of values for $\gamma$, considering
both weak and strong coupling between system and environment. For small values of $\gamma$,
the GP behaves very similarly to the unitary GP which is $\phi_U = \pi(1+\cos\theta_0)$. However,
as we increase the value of $\gamma$, there is a notable change in the curvature as a function of
$\theta_0$, leading to more values of $\theta_0$ with a null GP. As expected, the more decohering environment,
 the less survival of the GP.
In agreement with Fig.\ref{Fgchico}, we can see that the geometric phase is less corrected (with 
respect to the isolated case) 
in the presence of an ohmic non-equilibrium environment rather than of a supraohmic one. 
This can be noted by the fact that Fig.\ref{Fase3dn1} remains 
a smooth function of $\theta_0$  for bigger values of $\gamma$ than Fig.\ref{Fase3dn3}, in which case
the phase rapidly behaves different as a function of the dissipation constant.

The GP cannot be fully computed analytically but we can perform an
expansion in powers of the coupling constant, to obtain an accurate
approximation of it \cite{pra,nos,pau}. Hence, we expand in powers of $\gamma$  
the $\cos^2\theta_+$ in Eq.(\ref{fasecos}), using the definition
of the decoherence factors for each environment, namely Eqs.(\ref{fn1}) and (\ref{fn3}). As mentioned before, 
the correction to the GP is defined as $\delta \phi_G $, while $\phi_U$ is
the unitary GP.
In the case of the ohmic non-equilibrium environment, the correction to the unitary GP
is given by,
\begin{figure}[!ht]
\includegraphics[width=8cm]{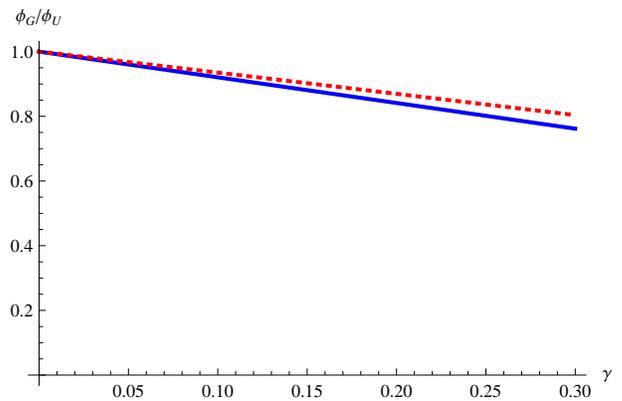}
\caption{(Color online) Comparison between exact GP (red dashed line), in the presence of an ohmic environment, and 
the first order perturbative correction (blue solid line) from Eq.(\ref{perturbohmic}). Perturbative result is in good agreement 
with the exact result for a long range of values of $\gamma$. Parameters used: $D/\Omega=1$, $\Omega\lambda = 1$, $\Lambda/\Omega = 1$.}
\label{perturb1}
\end{figure}

\begin{eqnarray}
 \delta \phi_{G_{n=1}} &\approx & \pi  \gamma \sin^2(\theta_0)
 \cos(\theta_0) \nonumber \\&+& \gamma \frac{\Omega D}{\Lambda^2} e^{-2 D \lambda} \sin^2(\theta_0)
 \cos(\theta_0).\label{perturbohmic}
\end{eqnarray}
 Proceeding the same way for the supraohmic environment, the correction to the unitary
GP is
\begin{eqnarray}
 \delta \phi_{G_{n=3}} & \approx & 6 \pi \gamma \sin^2(\theta_0)
 \cos(\theta_0) \nonumber \\
&+ & \gamma \frac{\Omega D^3}{4\Lambda^4} e^{-2 D \lambda} \sin^2(\theta_0)
 \cos(\theta_0). \label{perturbsupraohmic}
\end{eqnarray}

\begin{figure}[!ht]
\includegraphics[width=8cm]{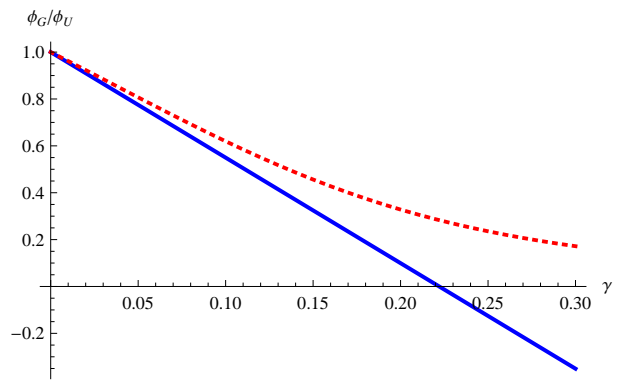}
\caption{(Color online) Comparison between exact geometric phase (red dashed line), in the presence of an supraohmic environment, and 
the first order perturbative correction (blue solid line) from Eq.(\ref{perturbsupraohmic}). Perturbative result is a good approximation for 
really small values of $\gamma$. Parameters used: $D/\Omega=1$, $\Omega\lambda = 1$, $\Lambda/\Omega = 1$ }
\label{perturb3}
\end{figure}

The corrections of the GP for both environments agree for small values of $\gamma$ in
Figs. \ref{Fase3dn1} and \ref{Fase3dn3}. In both cases, the dependence with the 
parameter $\lambda$ is exponentially negligible. 
Another interesting
feature of the corrections of the GP is that they depend
on the initial angle of the quantum state, and this dependence
is in agreement with the ones obtained for a two-level system 
in interaction with environments at equilibrium \cite{pra,nos,pau,pra2,pra3,prl}. Neglecting the 
small correction induced by $\lambda$ (which is a correct assumption seen Figs. \ref{Fase3dn1} and \ref{Fase3dn3}), both 
cases are similar to the very low temperature corrections found in \cite{pra} for 
the case of thermal environments. In Figs. \ref{perturb1} and \ref{perturb3}, we show the range of validity of 
the first order perturbative expansion in powers of $\gamma$. In Fig.\ref{perturb1}, it is clear that the 
perturbative result (solid line) of Eq.(\ref{perturbohmic}) is in excellent agreement with the exact result (dashed line), even for 
not too small values of the coupling 
strength parameter $\gamma$. Fig. \ref{perturb3} shows that Eq.(\ref{perturbsupraohmic}) is also a good approximation to 
the exact result (dashed line), but only for very small values of $\gamma$. 

\begin{figure}[!ht]
\includegraphics[width=10cm]{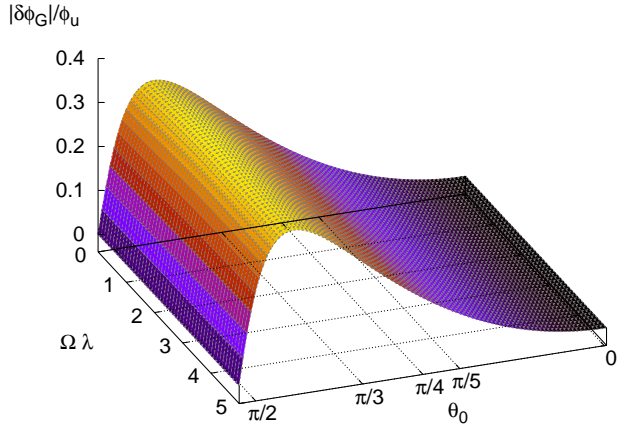}
\caption{(Color online) Behaviour of the geometric phase as a function of the initial state $\theta_0$ (in radians) of the
quantum system and the initial phases of the bath modes ($\lambda$) in a cycle
for an ohmic non-equilibrium environment.
Parameters used:  $\Lambda/\Omega= 1$; $\gamma = 3$; $D/\Omega = 0.1$}
\label{Fase3dn1lambda}
\end{figure}

It is an 
interesting feature to study which is the influence of the observed  dip (in the decoherence factor) 
in the behaviour of the GP. In this model of non-equilibrium environment, the parameter $\lambda$, 
which enters through the random phases,  sets the position at which the ``dip" or ``recoherence" takes place. 
 As we have seen, there is a time-scale when
the system seems to gain coherence ($t\sim \lambda$).  In Fig.\ref{Fase3dn1lambda},
we present the correction to the GP $|\delta \phi_G|$ as a function of the initial state of the quantum
system ($\theta_0$) and the initial phases of the bath modes ($\Omega \lambda $)for an ohmic non-equilibrium 
environment.
Therein, we can observe that the geometric phase is not affected by the dip in the
decoherence factor, as it is computed over a quasi-cyclic evolution. In Fig.\ref{Fase3dn1lambda}
we can note that the GP has a monotonous behaviour as a function of
$\lambda$. The analytical estimation of the influence 
of $\lambda$ in the correction to the GP, made in Eqs. (\ref{perturbohmic}) and (\ref{perturbsupraohmic}), 
is also checked in Fig. \ref{Fase3dn1lambda}.  

\section{Final Remarks}
\label{conclu}

The geometric phase of quantum states is an issue worth of attention.
It could be a potential application in holonomic quantum computation
since the study of spin systems effectively allows us to contemplate
the design of a solid state quantum computer. However, decoherence is the main
obstacle to overcome. All realistic quantum systems are coupled to their
surroundings to a greater or lesser extent. Furthermore, in most
cases of practical interest, quantum systems are subjected to many 
noise sources with different amplitudes and correlation times, corresponding
${\it de~facto}$ to a non-equilibrium environment.

Herein, we have presented a simple case to illustrate the general phenomenon of
dephasing in a non-equilibrium bath. We have studied the decoherence process
of a quantum system in interaction with an initially non-equilibrium bath that can be
controlled by manipulating the nature of the relative initial phases of the
bath modes. The decoherence factors computed here suggest that by engineering these
initial phases, the dephasing of the subsequent quantum evolution can potentially
be controlled. We have found similarities and differences of the decoherence
process between the environment presented here and thermal environments studied 
in previous works. 
The model presented here is another proposal for engineering
reservoirs in a manner reminiscent of a coherent control experiment using shaped pulses \cite{pulses}.  In this 
model, the control parameter $\lambda$ is derived not from a laser pulse, but rather from well-defined phase 
relations between the modes of the bath. Another possible candidate for realizing this decoherence environment, is to
use the artificially generated fluctuating environments with NMR. It could be possible, in principle, to use the quantum
simulator of Ref \cite{prl} to generate the fluctuating phase $\theta(\omega)$ of the
present proposal. 

The analysis of the effect produced by decoherence on the GP is 
crucial at the time to design an experimental setup to measure the GP using, for example, interferometry. We found 
that the convenient non-equilibrium environment to observe GPs is the weak coupled ohmic case. It is important 
that there is no restriction about zero temperature environments in this case, as it was found in \cite{pra}, as the 
most convenient scenario.
In this framework, these kind of environments could become
a proper experimental setup for the observation of the geometric phase.
Therefore, we have computed the geometric phase for an ohmic and supra-ohmic non-equilibrium
environment and seen how they deviate from the unitary geometric phase.
So far, we have seen that the characteristic dip of the decoherence factor
does not affect much the geometric phase and that ohmic non-equilibrium environment
are not as destructive as the supraohmic non-equilibrium ones or the thermal environments
we are so used to see in the literature. The effect done on
the geometric phase by the ohmic (or/and supraohmic) non-equilibrium environment
can be seen as similar to the one done by a single reservoir with an effective temperature, as the non-equilibrium environment 
model proposed in \cite{lutz}. 
In the very weak coupling limit, we have evaluated the corrections induced by the 
non-equilibrium environment on the unitary GP, showing that there is a small (exponentially suppressed) 
correction due to the random parameter $\lambda$. More general models should be 
analysed in a future work.\\

\acknowledgments

This work is supported by CONICET, UBA, and ANPCyT, Argentina.

\end{document}